\documentclass[a4paper,twocolumn,floatfix]{revtex4-2}

\usepackage{graphicx}
\usepackage[utf8]{inputenc}
\usepackage[T1]{fontenc}
\usepackage{amsmath,amssymb,amsfonts,amstext}
\usepackage{dsfont}
\usepackage[svgnames]{xcolor}
\usepackage{multirow}
\usepackage{enumitem}
\usepackage{threeparttable}
\usepackage{dcolumn}
\usepackage{booktabs}
\usepackage{caption}
\usepackage{amsmath}
\usepackage{subcaption} 
\begin{document}
\title{
Revealing evolution of Dark Energy density from observations}

\author{José Blanco$^{1,2}$}
    \email{jose.blancova@postgrado.uv.cl}
\author{V\'ictor H. C\'ardenas$^1$}
    \email{victor.cardenas@uv.cl}
\author{Cuauht\'emoc Campuzano $^2$}
    \email{ccampuzano@uv.mx}
    \affiliation{$^1$Universidad de Valparaíso, Instituto de Física y Astronomía, Av. Gran Bretana 1111, Valparaíso, Chile}
    \affiliation{$^2$Facultad de Física, Universidad Veracruzana, 91223 Xalapa-Enríquez, Ver., México }

\begin{abstract}
We present a model–independent reconstruction of the normalized dark energy density function, $X(z) \equiv \rho_{de}(z)/\rho_{de}(0)$, derived directly from the DES-SN5YR Type~Ia supernova sample. The analysis employs an inversion formalism that relates the derivative of the distance modulus, $\mu'(z)$, to the expansion history, allowing the data to determine the shape of $X(z)$ without assuming a specific equation–of–state or dark energy density parameterization.  
A statistically optimized binning of the supernova sample (using 17 intervals following the Freedman–Diaconis criterion and 34 following Scott’s rule) ensures a stable estimation of $\mu'(z)$ and a controlled propagation of uncertainties throughout the inversion process.
The resulting $X(z)$ remains statistically consistent with a constant value within one standard deviation across the entire redshift range, showing no significant evidence for an evolving dark energy component at present. 
In a direct comparison among $\Lambda$CDM, CPL, and the quadratic $X^2(z)$ parameterization —where CPL and $X^2(z)$ each introduce two additional free parameters relative to $\Lambda$CDM— the CPL model attains the best statistical agreement with the data, albeit only marginally and strictly within this restricted model set. 
These outcomes indicate that current observations are compatible with an almost constant dark energy density ($w \simeq -1$), while the inversion framework remains sensitive to subtle departures that forthcoming high-precision surveys could resolve.
\end{abstract}

\date{\today} 

\maketitle

\section{Introduction}

The nature of dark energy (DE) remains one of the most profound open problems in modern cosmology and fundamental physics \cite{Frieman:2008sn, Weinberg:2013agg}. Dark energy is invoked to explain the accelerated expansion of the Universe, first revealed through observations of Type~Ia supernovae nearly three decades ago \cite{SupernovaCosmologyProject:1998vns, SupernovaSearchTeam:1998fmf}. The simplest theoretical interpretation corresponds to a cosmological constant $\Lambda$, leading to the highly successful $\Lambda$CDM framework. Despite its empirical success, this model faces deep conceptual issues, including the cosmological constant problem, the fine-tuning and coincidence problems, and emerging observational tensions in $H_0$ and $\sigma_8$ determinations \cite{Blum:2024igb, DES:2024wym}.

Several theoretical extensions have been proposed to alleviate these issues. These can broadly be classified into: (i) models of \textit{dynamical dark energy}, such as quintessence, phantom, or k-essence fields \cite{Chaudhary:2025bf, Malekjani:2023ple, DiValentino:2025otz}; (ii) \textit{modified gravity} frameworks, which alter the Einstein–Hilbert action, including $f(R)$, $f(T)$, and scalar–tensor theories \cite{Sahlu:2025jeh, Paliathanasis:2025hjw, Samaddar:2025dyj, Ishak:2024jhs}; and (iii) models that relax the Copernican principle or allow inhomogeneous geometries \cite{Giare:2024oil}.  
Each class introduces different phenomenological imprints and potential solutions to current cosmological tensions \cite{Afzal:2024foz, Chen:2024vuf}. Comprehensive taxonomies and comparative studies can be found in \cite{Gialamas:2024lyw, Colgain:2021pmf, Lee:2025pzo}.

Parallel to theoretical model building, the cosmological community has devoted significant effort to reconstructing the behavior of dark energy directly from observational data, with minimal theoretical assumptions. This approach aims to extract the redshift evolution of key cosmological quantities—such as the equation of state $w(z)$, the Hubble rate $H(z)$, or the normalized density $X(z) = \rho_{\rm de}(z)/\rho_{\rm de}(0)$—using either parametric or nonparametric techniques \cite{Gialamas:2025pwv, Li:2025ula, KiDS:2020suj}.  
This “data-driven” or “agnostic” strategy has become increasingly attractive given the precision of recent late-time surveys such as DES-SN5YR and DESI (DR1 and DR2), which jointly probe the expansion history with unprecedented accuracy \cite{Li:2025muv,Li:2025ops,DESI:2024kob,Giare:2024smz}.

Among parametric models, the Chevallier--Polarski--Linder (CPL) parametrization 
\cite{Chevallier:2000qy, Linder:2002et} defines the dark-energy equation of state as
\begin{equation}
    w(z) = w_0 + w_a \frac{z}{1+z},
\end{equation}
and has long served as a benchmark for exploring dynamical dark energy.
Its analytic simplicity and two-parameter form make it highly tractable for data analysis 
\cite{Shajib:2025tpd, Arjona:2024dsr}.
However, the CPL model's rigidity constrains the shape of $w(z)$, preventing the modeling 
of negative effective densities or sharp transitions across $w=-1$ 
\cite{Alam:2025epg, Malekjani:2023ple}.

A quadratic parametrization of $X(z)$, presented in the literature as the $X^2(z)$ model 
\cite{Cardenas:2014jya, CosmoVerseNetwork:2025alb, Cardenas:2014jya}, allows a broader phenomenological range, 
including curvature or sign changes in the reconstructed density.
This flexibility, however, increases degeneracies and may amplify statistical noise 
\cite{Grandon:2021nls, Li:2025muv}.
The present work builds on these developments, employing a quadratic form to test for 
possible low-$z$ deviations from $\Lambda$CDM using homogeneous DES-SN5YR data.

Beyond fixed parametrizations, nonparametric and hybrid reconstructions have gained prominence. 
Bayesian and Gaussian Process regressions have been used to infer $w(z)$ without prescribing its shape \cite{Rajvanshi:2019wmw, KiDS:2020suj}, 
while machine learning and symbolic regression frameworks \cite{Sousa-Neto:2025gpj, Mitra:2024ahj} explore model-independent representations of cosmic expansion. 
Such techniques are particularly suited to identifying small departures from $\Lambda$CDM suggested by DESI BAO measurements \cite{Ishak:2024jhs, Park:2024vrw}. 
In parallel, several works have employed binned and spline-based approaches to reconstruct the luminosity-distance relation of Type Ia supernovae 
and its redshift dependence \cite{Shajib:2025tpd, Giare:2024smz, Li:2024qso}, 
providing complementary, data-driven probes of late-time cosmic acceleration.

Recently, analyses of DESI (DR1 and DR2) and DES-SN5YR have reignited the debate over whether dark energy evolves with time. Dynamical models \cite{Yang:2025mws, Li:2025ops, Blum:2024igb} and interacting scenarios \cite{Giare:2024smz, Li:2024qso} appear slightly favored over $\Lambda$CDM in some redshift bins, though the statistical significance remains modest. Modified gravity interpretations \cite{Sahlu:2025jeh, Paliathanasis:2025hjw} and local inhomogeneity effects \cite{Gialamas:2024lyw} have also been explored as alternative explanations. Still, others argue that when systematic effects and prior assumptions are rigorously marginalized, the data remain statistically consistent with a cosmological constant \cite{Giare:2024oil, Park:2024vrw, DESI:2024kob}.

In light of this evolving landscape, a reconstruction approach that directly connects data to the dark energy density function $X(z)$—without imposing strong parametric priors—is particularly valuable. The \textit{inversion method} pursued in this work achieves this by deriving $X(z)$ directly from the observed $\mu(z)$ relation. This framework minimizes theoretical bias and provides a natural avenue for testing the internal consistency of DES-SN5YR data and for comparing with both CPL and quadratic parametrizations. It also enables a clearer quantification of how observational uncertainties and binning strategies propagate into the reconstructed density.

The rest of this paper is organized as follows. Section~\ref{sec:data} describes the observational datasets employed, including the DES-SN5YR supernova sample, DESI BAO measurements, and redshift-space distortions (RSD). Section~\ref{sec:3} outlines the reconstruction and inversion methodologies used to derive the derivative of the distance modulus, $\mu'(z)$, and to infer the normalized dark energy density $X(z)$. Section~\ref{sec:4} presents the statistical treatment of the binned data and the construction of $\mu'(z)$. Section~\ref{sec:5} reports the reconstructed functions and compares them with the $\Lambda$CDM, CPL, and quadratic $X^2(z)$ parameterizations. Section~\ref{sec:6} discusses the cosmological implications and robustness of the results, while Section~\ref{sec:7} summarizes our conclusions and future perspectives.

\section{Data}
\label{sec:data}

\subsection{Type Ia Supernovae: DES-SNY5R}

The principal dataset employed in our analysis is the DES 5YR (DES-SN5YR) Type Ia supernova (SN~Ia) sample, recently released by the Dark Energy Survey Collaboration \cite{DES:2025tir, DES:2024ffp}. This compilation constitutes the final cosmology-grade data set from five years of DES observations with the DECam instrument at the Blanco 4m telescope in Cerro Tololo. It includes approximately $1500$ SNe~Ia spanning the redshift range $0.01 \lesssim z \lesssim 1.2$, making it the largest homogeneous SN~Ia sample from a single survey to date.  

Assuming a flat universe, distances are estimated using the luminosity distance relation,
\begin{equation}
    D_L(z) = c \,(1+z)\int_0^z \frac{dz'}{H(z')},
\end{equation}
which enters the expression for the apparent magnitude,
\begin{equation}
    m(z) = 5 \log_{10}\!\left(\frac{D_L(z)}{h}\right) + M,
\end{equation}
where $M$ denotes the absolute magnitude of SNe~Ia. Cosmological parameters governing $H(z)$ are then fit by comparing the theoretical prediction in Eq.~(2) with the observed light-curve calibrated magnitudes from the DES-SN5YR catalog.  

Compared to heterogeneous compilations such as Pantheon+, the DES-SN5YR dataset offers key advantages: the use of a single instrument ensures uniform photometry, while systematic uncertainties are carefully modeled, including photometric calibration, Malmquist bias, host-galaxy correlations, and dust extinction \cite{DES:2024jxu, DES:2021vln}. These improvements significantly reduce survey-to-survey systematics and provide a robust foundation for cosmological inference. 
In this work, we adopt the full DES-SN5YR sample without imposing external priors on $H_0$, in line with our agnostic reconstruction strategy that lets the data determine the shape of $X(z)$.


\subsection{Baryon Acoustic Oscillations: DESI DR1 \& DR2}

The Dark Energy Spectroscopic Instrument (DESI) has set a new standard in the precision mapping of the large-scale structure of the Universe. 
Its first data release (DR1) delivered robust baryon acoustic oscillation (BAO) measurements across multiple tracers, including the Bright Galaxy Sample (BGS), Luminous Red Galaxies (LRG), Emission Line Galaxies (ELG), and Quasars (QSO), covering the range \(0.1 \lesssim z \lesssim 2.1\) \cite{DESI:2024mwx,DESI:2025zgx}. 
The second data release (DR2), comprising the first three years of observations, expanded this sample to more than 14 million objects, providing the most comprehensive spectroscopic mapping of the cosmic web to date. 

BAO positions are determined both in the transverse and radial directions, defined respectively by
\begin{equation}
D_M(z) = c \int_0^z \frac{dz'}{H(z')}, \qquad 
D_H(z) = \frac{c}{H(z)} ,
\end{equation}
while for lower signal-to-noise measurements an isotropic volume-averaged distance is used,
\begin{equation}
D_V(z) = \left[ z D_M^2(z) D_H(z) \right]^{1/3}.
\end{equation}
All distances are expressed in units of the comoving sound horizon at the baryon-drag epoch, \(r_d\). 
Following its integral definition,
\begin{equation}
r_d = \int_{z_d}^{\infty} \frac{c_s(z)}{H(z)} \, dz ,
\end{equation}
with \(c_s(z) = c / \sqrt{1 + 3\rho_b(z)/(4\rho_\gamma(z))}\), the calculation ensures consistency between early- and late-time quantities without adopting external priors. 

The precision achieved by DESI DR2 in measuring \(D_H(z)/r_d\) and \(D_M(z)/r_d\) now exceeds that of all previous spectroscopic surveys, providing a fundamental anchor for testing potential deviations from the \(\Lambda\)CDM expansion. 
When combined with homogeneous supernova samples such as DES-SN5YR, these data enable a direct reconstruction of the dark energy density function \(X(z)\) without imposing strong external priors on \(H_0\) or \(\Omega_m\).


\subsection{Redshift-Space Distortions (RSD)}

Beyond the geometrical information provided by BAO, redshift-space distortions (RSD) trace the growth of cosmic structure through anisotropies in the clustering of galaxies. 
The observable parameter
\begin{equation}
f\sigma_8(z) = f(z) \, \sigma_8(z),
\end{equation}
combines the linear growth rate \(f = d\ln D / d\ln a\) with the amplitude of matter fluctuations \(\sigma_8\), offering a direct test of the gravitational framework at late times \cite{Kazantzidis:2019nuh,Adil:2023jtu}. 

In the linear regime, the evolution of matter perturbations is governed by
\begin{equation}
\delta''(a) + \left[ \frac{3}{a} + \frac{E'(a)}{E(a)} \right] \delta'(a)
- \frac{3}{2} \frac{\Omega_{m0}}{a^5 E^2(a)} \, \delta(a) = 0 ,
\end{equation}
where \(E(a) = H(a)/H_0\). 
Solving this equation allows the theoretical prediction of \(f\sigma_8(z)\), which can be compared directly with DESI RSD measurements and independent datasets \cite{Percival:2008sh,Gannouji:2018ncm}. 

While individual RSD measurements are less precise than BAO distances, their joint analysis provides essential dynamical constraints: BAO characterize the background geometry, whereas RSD probe first-order perturbation dynamics. 


\section{Methods}
\label{sec:3}

In this section, we will review two methods for extracting information from observational data, aiming to provide insights into the behavior of dark energy as a function of redshift. In the first subsection, we will examine the case of parametric reconstruction, focusing on two approaches discussed in the literature. Then, in a second subsection we introduce the inversion method, and present the corresponding results.

\subsection{Reconstruction Method}

{\color{blue} 
}

The reconstruction method consists in selecting a cosmological ansatz function that depends on a small set of free parameters, which are then constrained using statistical tools such as Markov Chain Monte Carlo (MCMC). The idea is to use this function as a probe of eventual variation with cosmological redshift. In this work, we focus on the normalized dark energy density, defined as
\begin{equation}
    X(z) \equiv \frac{\rho_x(z)}{\rho_x(0)}.
\end{equation}

\noindent
 
A widely used ansatz is the Chevallier–Polarski–Linder (CPL) parameterization \cite{Chevallier:2000qy,Linder:2002et}, which assumes a time-varying equation of state for dark energy,
\begin{equation}
    w(z) = w_0 + \frac{w_a z}{1+z},
\end{equation}
where $w_0$ and $w_a$ are free parameters. From this, the normalized DE density takes the form
\begin{equation}\label{eq:Xcpl}
    X_{\rm CPL}(z) = (1+z)^{3(1+w_0+w_a)} \exp\!\left(-3\frac{w_a z}{1+z}\right).
\end{equation}
The $\Lambda$CDM model is recovered in the special case $w_0=-1$, $w_a=0$. As pointed out in the introduction, this parametrization is restricted by construction to yield positive values of $X(z)$.

\noindent
 
Alternatively, following recent works \cite{Cardenas:2014jya,Bernardo:2021cxi,Orchard:2024bve,Bernardo:2022pyz}, one may directly parametrize $X(z)$ using a quadratic function,
\begin{equation}\label{eq:Xquad}
    X_{\rm q}(z) = 1 + \frac{z(4x_1 - x_2 - 3)}{z_m} - \frac{2z^2(2x_1 - x_2 - 1)}{z_m^2},
\end{equation}
where $z_m$ is the maximum redshift of the dataset, and the free parameters are $x_1 = X(z_m/2)$ and $x_2 = X(z_m)$. The $\Lambda$CDM case corresponds to $x_1 = x_2 = 1$. Unlike the CPL form, this quadratic parametrization allows $X(z)$ to take negative values, as discussed in the introduction.

Both parametrizations introduce two free parameters that can be directly constrained by observational data. The CPL form is the historical standard, but structurally limited to positive dark energy densities, while the quadratic form is more flexible, at the cost of potentially introducing features that reflect the choice of functional form rather than genuine cosmological trends. This complementarity highlights the motivation to explore different approaches within the reconstruction program.

\subsection{Inversion Method}

Rather than relying on the specific form of the parametrization for the probe function, 
we can instead directly extract the implied value of \( X(z) \) from the observational data. 
In this approach, we simply use the relations that connect the observational data 
at each redshift \( z_i \) with the corresponding dark energy density \( X(z_i) \).

To illustrate the method, we use type Ia supernova data as an example. In this case, the observational quantity is the distance modulus defined theoretically by
\begin{equation}\label{eq: mu_theo}
    \mu_B(z) = 5 \log_{\text{10}} \left[\frac{d_L(z)}{\text{10 pc}} \right],
\end{equation}
where the luminosity distance $d_L(z)$ depends on the cosmological model through
\begin{equation}\label{eq: dl_def}
    d_L(z) = \frac{c}{H_0} (1+z) \int_{0}^{z} \frac{dz'}{E(z')},
\end{equation}
assuming a flat space, with $E(z)=H(z)/H_0$. We use a spatially flat Friedmann–Lemaître–Robertson–Walker (FLRW) geometry, a result in agreement with current constraints from CMB anisotropies and BAO data, which consistently indicate $|\Omega_k| < 10^{-3}$ \cite{Planck:2018vyg,DESI:2023ytc}. 
The total energy content of the Universe is assumed to be dominated at late times by two components: a pressureless matter sector, representing both baryons and cold dark matter, evolving as $(1+z)^3$, and a dark energy component described by the normalized density function $X(z) = \rho_{\rm de}(z)/\rho_{\rm de}(0)$. Under these assumptions, the expansion rate is given by
\begin{equation}
    H^2(z) = H_0^2\left[\Omega_m(1+z)^3 + (1 - \Omega_m) X(z)\right].
\end{equation}

Observationally, the combination of Planck CMB measurements, DESI BAO distances, and DES–SN5YR supernovae strongly constrains spatial curvature, 
while the contribution from radiation and relativistic species becomes negligible at $z \lesssim 2$. The aim is to reverse the relationship so that, for a given observational value $\mu (z)$, we can determine the corresponding point $X(z)$. Starting from equation (\ref{eq: mu_theo}) and explicitly incorporating the units for $c/H_0$, we have

\begin{equation}
    10^{(\mu_B - 25)/5} = \frac{c}{H_0}\left[\int_0^{z} \frac{dx}{E(x)} + \frac{1+z}{E(z)} \right],
\end{equation}
from which we can differentiate with respect to the redshift $z$ to obtain
\begin{equation}\label{eq: key}
X(z) = \frac{1}{1-\Omega_m} \left[ -\Omega_m (1+z)^3 + B^{-2}(z) \right], 
\end{equation}
where
\begin{equation}\label{eq: keyeq}
    B(z) = 10^{(\mu_B - 25 )/5}\left\{ \frac{\ln(10)\mu'_B(z)}{5(1+z)} - \frac{1}{(1+z)^2} \right\}.
\end{equation}
The inverted equation depends on the derivative of the apparent magnitude $\mu'_B(z)$, which must be derived from the data points of $\mu_B(z)$. In the following analysis, we assume the best-fit values for $\Omega_m$ and $H_0$ from two different sources: Planck \cite{Planck:2018bsf} and SH0ES \cite{Riess:2021jrx}. We use the DES5Y data for type Ia supernova.

In the analyses that follow, the parametric reconstructions of $X(z)$---namely the $\Lambda$CDM, CPL, and quadratic $X_2(z)$ models---are constrained using the joint combination of DES--SN5YR supernovae, DESI BAO distances, and redshift–space–distortion (RSD) measurements. 
This multi–probe configuration ensures consistency between background expansion and structure–growth observables, providing the most reliable assessment of cosmological parameters. 
By contrast, the inversion procedure introduced above is applied \emph{exclusively} to the DES--SN5YR Type~Ia supernova sample, from which the non–parametric $X(z)$ is obtained directly from the luminosity–distance relation.

\section{Construction of the derivative of $\mu$}
\label{sec:4}
The analysis of the relationship between the distance modulus $\mu$ and the redshift $z$ requires an appropriate discretization of the observational domain, in order to preserve the statistical structure of the dataset without introducing spurious noise. To this end, a uniform binning procedure in $z$ was applied.

\subsection{Statistical Criteria for the Choice of the Number of Bins}

The selection of the number of bins, $N_{\text{bins}}$, is a classical problem in statistics, as it governs the trade-off between resolution and noise. A small number of bins may conceal relevant variations, whereas an excessive number amplifies measurement noise. To determine $N_{\text{bins}}$ objectively, two well-established rules discussed in the literature were employed: the Scott’s rule \citep{10.1093/biomet/66.3.605,https://doi.org/10.1002/wics.118} and the Freedman–Diaconis rule \citep{https://doi.org/10.1007/BF01025868}.

Both methodologies define the optimal bin width, $h$, as a function of the total number of observations $N$ and a statistical measure of dispersion. In Scott’s rule, a nearly Gaussian distribution is assumed, and the width is computed as:
\begin{equation}
h_{\text{Scott}} = 3.49 \sigma N^{-1/3},
\end{equation}
where $\sigma$ is the sample standard deviation. In contrast, the Freedman–Diaconis rule employs the interquartile range,
\begin{equation}
\mathrm{IQR} = Q_3 - Q_1,
\end{equation}
which is tends to be more robust against non-normal distributions, defining:
\begin{equation}
h_{\text{FD}} = 2, \mathrm{IQR}, N^{-1/3}.
\end{equation}

The number of bins is then obtained as:
\begin{equation}
N_{\text{bins}} = \frac{z_{\max} - z_{\min}}{h}.
\end{equation}

Thus, the Freedman–Diaconis method produces a more stable partition, less sensitive to asymmetric dispersion, whereas the Scott method provides finer resolution at the expense of greater susceptibility to noise. In the present work, both schemes were adopted complementarily: ($N_{\text{bins}} = 17$)  and $N_{\text{bins}} = 34$ to assess the stability and reproducibility of the local trend of $\mu(z)$.

\subsection{Construction of the Binned Magnitude and Its Derivative}

The redshift range \([z_{\min}, z_{\max}]\) was divided into \(N_{\text{bins}}\) equally spaced intervals.  
For each bin \(i\), the central redshift was defined as
\begin{equation}
    \bar{z}_i = \frac{z_i + z_{i+1}}{2},
\end{equation}
where \(z_i\) and \(z_{i+1}\) correspond to the lower and upper limits of the interval, respectively.  
The mean distance modulus within each bin was computed as
\begin{equation}
    \bar{\mu}_i = \frac{1}{N_i} \sum_{j=1}^{N_i} \mu_j,
\end{equation}
where \(N_i\) is the number of supernovae contained in bin \(i\).  
The corresponding uncertainty was estimated from the sample variance as
\begin{equation}
    \sigma_{\bar{\mu},i} = 
    \sqrt{\frac{1}{N_i - 1} \sum_{j=1}^{N_i} \left(\mu_j - \bar{\mu}_i\right)^2}.
\end{equation}
Alternatively, the standard error on the mean can be obtained as 
\(\sigma_{\bar{\mu},i} / \sqrt{N_i}\) if required for subsequent analyses.


To estimate the variation of magnitude with redshift, a continuous function $\mu(z)$ was interpolated using a cubic spline fitted to the binned values, and the derivative was evaluated analytically as:
\begin{equation}
\mu'(z) = \frac{d\mu}{dz}.
\end{equation}
The uncertainty associated with the derivative was obtained via a Monte Carlo procedure, generating perturbed realizations of $\bar{\mu}*i$ within their respective error bars $\sigma*{\bar{\mu},i}$, and recalculating the spline at each iteration. The resulting dispersion of $\mu'(z)$ values defines the uncertainty:
\begin{equation}
\sigma_{\mu'(z_i)} = \sqrt{ \langle \mu'(z_i)^2 \rangle - \langle \mu'(z_i) \rangle^2 }.
\end{equation}

This procedure allows simultaneous quantification of the mean behavior $\bar{\mu}(z)$ and its local derivative $\mu'(z)$, while minimizing the influence of observations with excessive errors and ensuring the statistical stability of the reconstructed curves.

\section{Results}
\label{sec:5}

\begin{table*}[htbp]
    \centering
    \resizebox{0.95\textwidth}{!}{%
    \begin{tabular}{lcccccccc}
        \toprule
        & $\Omega_m$ & $H_0r_d$ & $\sigma_8$ & $w_0/x_1$ & $w_a/x_2$ & $\chi^2_\text{red.}$ & $\Delta$AIC & $\Delta$BIC \\
        \midrule
        \textbf{Flat $\Lambda$CDM} \\
        DES & $0.314 \pm 0.025$ & $101.1 \pm 2.0$ & $0.758 \pm 0.031$ & $-1$ & $0$ & $0.96$ & $0$ & $0$ \\
        DES+DESI+RSD & $0.317 \pm 0.022$ & $100.5 \pm 1.9$ & $0.754 \pm 0.029$ & $-1$ & $0$ & $0.93$ & $0$ & $0$ \\
        \midrule
        \textbf{Flat CPL} \\
        DES & $0.328 \pm 0.028$ & $101.8 \pm 2.1$ & $0.746 \pm 0.030$ & $-0.95 \pm 0.12$ & $-0.68 \pm 0.25$ & $0.94$ & $+2.1$ & $+4.7$ \\
        DES+DESI+RSD & $0.325 \pm 0.019$ & $101.2 \pm 1.8$ & $0.752 \pm 0.027$ & $-0.97 \pm 0.09$ & $-0.62 \pm 0.20$ & $0.92$ & $+1.8$ & $+4.2$ \\
        \midrule
        \textbf{Flat $X^2(z)$CDM} \\
        DES & $0.335 \pm 0.020$ & $102.3 \pm 2.0$ & $0.730 \pm 0.028$ & $0.42 \pm 0.14$ & $-1.74 \pm 0.60$ & $0.91$ & $+2.4$ & $+5.0$ \\
        DES+DESI+RSD & $0.336 \pm 0.013$ & $102.1 \pm 1.7$ & $0.731 \pm 0.027$ & $0.40 \pm 0.13$ & $-1.69 \pm 0.48$ & $0.90$ & $+1.9$ & $+4.3$ \\
        \bottomrule
    \end{tabular}
   }
    \caption{Best-fit cosmological parameters for the flat $\Lambda$CDM, CPL and $X2$CDM models using DES-SN5YR and DESI+DES+RSD data combinations. The last two columns report the relative Akaike (AIC) and Bayesian (BIC) information criteria with respect to the $\Lambda$CDM baseline model.}
\end{table*}
\begin{table}[htbp]
\centering
\scriptsize
\resizebox{0.44\textwidth}{!}{%
\begin{tabular}{lc}
\toprule
\textbf{Parameter} & \textbf{Prior} \\
\midrule
$\omega_b$ & $0.02218 \pm 0.00055$ \\

\midrule
$\Omega_m$ & $[0,1]$ \\
$H_0r_d$ [km\,s$^{-1}$\,Mpc$^{-1}$] & $[90,120]$ \\
$\sigma_8$ & $[0.5,1.1]$ \\
$w_0$ & $[-2,0]$ \\
$w_a$ & $[-2,2]$ \\
$x_1$ & $[-2,2]$ \\
$x_2$ & $[-3,1]$ \\
\bottomrule
\end{tabular}
}
\caption{Prior ranges adopted for all MCMC analyses. 
Upper block: Gaussian priors from early-time physics (BBN and CMB). 
Lower block: flat, non-informative priors used in $\Lambda$CDM, CPL, and $X2$CDM runs.}
\end{table}

\begin{figure}[htbp]
    \centering
    \begin{subfigure}{\columnwidth}
        \centering
        \includegraphics[width=\columnwidth]{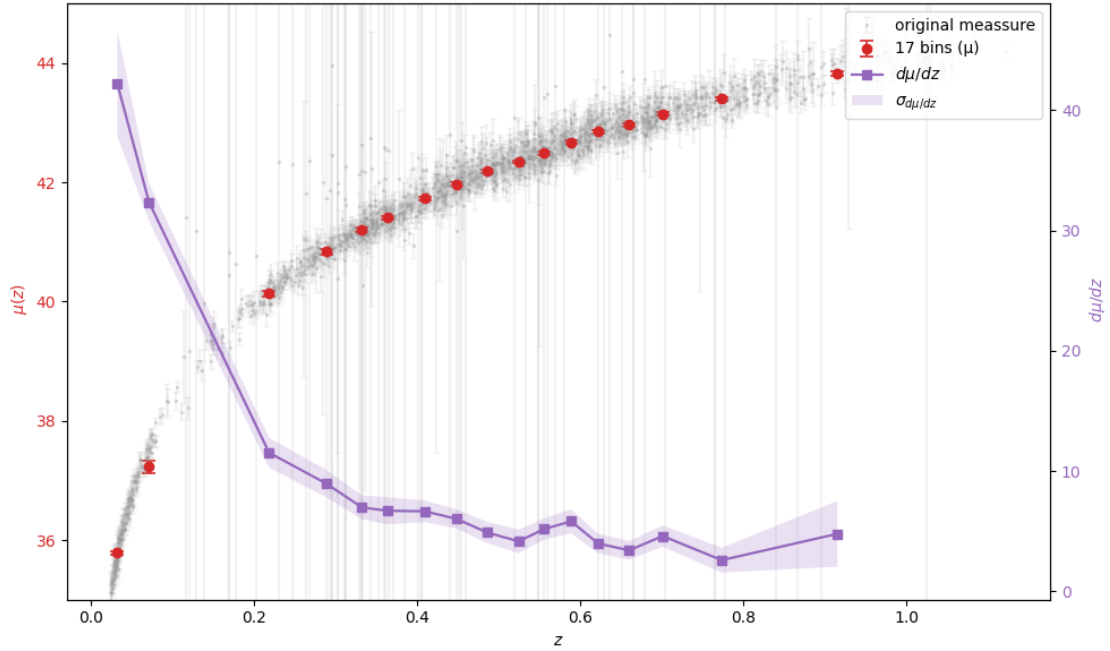}
        \caption{Reconstruction of the derivative of the distance modulus $\mu'(z)$ using 17 redshift bins following the Freedman–Diaconis criterion.}
        \label{fig:subfig1}
    \end{subfigure}

    \vspace{0.4cm} 

    \begin{subfigure}{\columnwidth}
        \centering
        \includegraphics[width=\columnwidth]{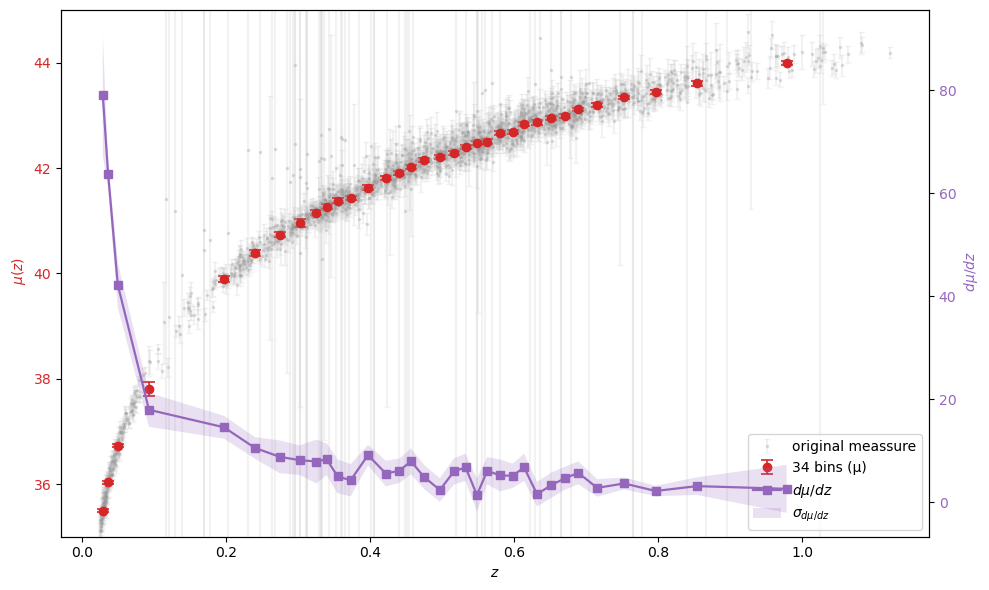}
        \caption{Same as the top panel but using 34 redshift bins determined from Scott’s rule. }
        \label{fig:subfig2}
    \end{subfigure}

    \caption{Derivative of the distance modulus $\mu'(z)$ reconstructed from the DES-SN5YR supernova sample. The smooth black curve represents the spline interpolation, while the shaded band indicates the $1\sigma$ uncertainty derived from Monte Carlo realizations.}
    \label{fig:one_vertical}
\end{figure}

Figure~\ref{fig:one_vertical} shows the reconstruction of the derivative of the distance modulus, $\mu'(z)$, obtained from the DES-SN5YR supernova sample. The reconstruction yields a smooth, monotonic behavior over the full range of observations. At low redshift ($z\lesssim0.2$), $\mu'(z)$ increases nearly linearly, consistent with the homogeneous expansion in the local universe. At intermediate redshifts ($0.3\lesssim z\lesssim0.6$), the slope gradually decreases, indicating the transition toward accelerated expansion. Beyond $z\simeq0.8$, the derivative becomes noisier due to the reduced number of supernovae per bin, yet no discontinuities or oscillatory artifacts are observed. The shape of the reconstructed function demonstrates that the derivative procedure preserves the intrinsic curvature of $\mu(z)$ and that statistical noise remains subdominant.

\begin{figure}[htbp]
    \centering
    \begin{subfigure}{\columnwidth}
        \centering
        \includegraphics[width=\columnwidth]{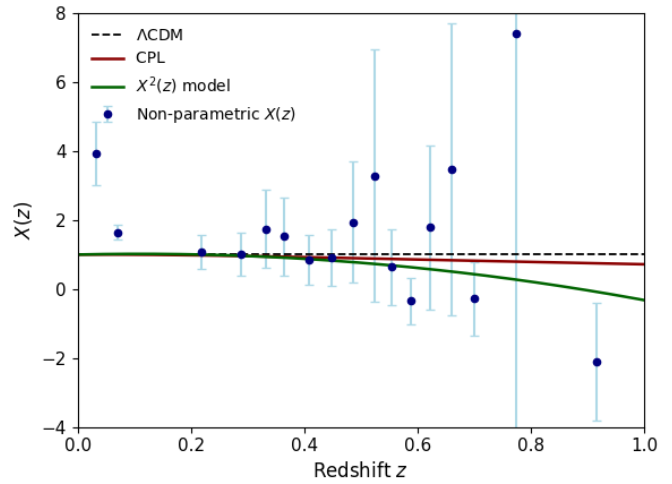}
        \caption{Reconstructed dark energy density function $X(z)$ obtained from the 17-bin (Freedman–Diaconis) derivative.}
        \label{fig:subfig3}
    \end{subfigure}

    \vspace{0.4cm} 

    \begin{subfigure}{\columnwidth}
        \centering
        \includegraphics[width=\columnwidth]{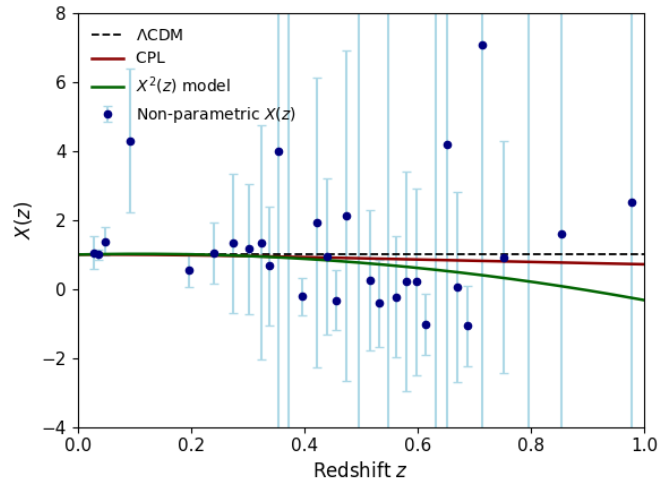}
        \caption{Reconstructed $X(z)$ using the 34-bin (Scott) derivative.}
        \label{fig:subfig4}
    \end{subfigure}

    \caption{Comparison of the reconstructed dark energy density $X(z)$ using two statistically motivated binning schemes.  Blue points with light–blue error bars show the non–parametric inversion; the black dashed line represents $\Lambda$CDM, and the red and green curves correspond to the CPL and quadratic $X^2(z)$ parameterizations, respectively.}
    \label{fig:two_vertical}
\end{figure}

Applying the inversion method to the reconstructed $\mu'(z)$ yields the dark energy density function $X(z)$ shown in Fig.~\ref{fig:two_vertical}. The non-parametric inversion is compared with two reference parametrizations: the Chevallier–Polarski–Linder (CPL) form and the quadratic model $X^2(z)$. The reconstructed $X(z)$ remains statistically consistent with unity up to $z\simeq0.5$, showing no indication of a time-varying dark energy component. A mild curvature appears at higher redshift, which the quadratic model reproduces naturally, whereas the CPL form (restricted to a monotonic evolution) cannot fully capture it.

The resulting $X(z)$ reproduces the constant behavior expected for a cosmological constant, while the small curvature observed at intermediate redshift remains within statistical uncertainty. The agreement between parametric and non-parametric reconstructions confirms the robustness of the method.

\section{Discussion}
\label{sec:6}

The reconstructed dark energy density obtained through the inversion of $\mu'(z)$ provides an empirical, model–independent assessment of late–time cosmic acceleration. The near–constancy of $X(z)$ found here agrees with the most recent large–scale analyses of DESI and DES-SN5YR data, which consistently indicate that any deviation from a cosmological constant remains below statistical significance \cite{Li:2025ops, Blum:2024igb, Yang:2025mws, Malekjani:2023ple}. Our reconstruction therefore supports the view that the apparent tensions in the late–time expansion rate—such as the differences in $H_0$ or $\sigma_8$—do not necessarily imply dynamical dark energy but may instead originate from residual systematics or local–structure effects \cite{Gialamas:2024lyw, Sahlu:2025jeh}.

From a methodological perspective, the inversion technique isolates the information content of the luminosity–distance relation more transparently than conventional parameter fits. By focusing on $\mu'(z)$, the method exposes the statistical leverage of each redshift interval and makes explicit the propagation of observational uncertainties to $X(z)$.  
The absence of oscillatory or unstable behavior in the reconstructed derivative confirms that the smoothing and differentiation procedures preserve the intrinsic curvature of the data.  
This stability is particularly relevant when compared with Gaussian–process or spline–based reconstructions, which can introduce artificial inflection points if the regularization or kernel width is not physically motivated \cite{KiDS:2020suj, Rajvanshi:2019wmw}.  

The mild curvature observed at $z \gtrsim 0.5$—well described by the quadratic model—illustrates the importance of flexible but controlled parametrizations.  
Quadratic representations provide a minimal extension to $\Lambda$CDM capable of accommodating weak trends without overfitting noise, unlike higher–order polynomial or piecewise models that tend to inflate uncertainties \cite{Grandon:2021nls}.  
The fact that the curvature remains within one standard deviation indicates that present supernova data alone cannot establish any time dependence in the dark energy density.  
Nevertheless, similar features have been reported in DESI–based reconstructions \cite{Ishak:2024jhs, Li:2025muv}, where the inferred $w(z)$ slightly departs from $-1$ at intermediate redshift.  
The recurrence of such mild trends across independent analyses warrants attention, though current data do not yet allow a distinction between a physical signal and statistical covariance among the redshift bins.

\noindent
To quantify the impact of background parameters on the reconstruction, we tested the models under variations of both the Hubble constant and the present matter density. Changing $H_0$ from $67.4$ to $73.0~\mathrm{km\,s^{-1}\,Mpc^{-1}}$ produces modest shifts in the statistical distance of all models, with $N_\sigma(\Lambda\mathrm{CDM})$ ranging from $1.50$ to $0.36$ and $N_\sigma(\mathrm{CPL})$ from $1.49$ to $0.36$, while $X^2(z)$ remains within $1.34 > N_\sigma > 0.48$. Varying $\Omega_M$ within $0.28{-}0.34$ yields comparable effects, with changes in $N_\sigma$ below $0.3$ across all cases. These small shifts confirm that the reconstructed $X(z)$ is stable under 

modifications of the background expansion. Among the three parameterisations, the CPL form consistently achieves the lowest $\chi^2$ and $N_\sigma$ values, typically $N_\sigma \simeq 1.2{-}1.3$ for the baseline configuration ($H_0=70$, $\Omega_M=0.31$), indicating the best statistical compatibility with the reconstructed dark energy density. The $\Lambda$CDM and quadratic models reproduce the same general curvature but with slightly higher residuals, suggesting that a weakly dynamical evolution of dark energy---captured by the CPL form---offers the most faithful description of the data.

Possible interpretations of these subtle features range from genuine dynamical dark energy to apparent evolution induced by local inhomogeneities or backreaction effects.  
Analyses of large–scale voids and anisotropic clustering suggest that local density contrasts can mimic an effective evolution in $X(z)$ when averaged over limited volumes \cite{Gialamas:2024lyw}.  
Alternatively, late–time modified–gravity scenarios, including non–minimally coupled scalar–tensor models, can generate similar shallow curvature while remaining consistent with other cosmological probes \cite{Paliathanasis:2025hjw, Samaddar:2025dyj}.  
Although our results do not favor such mechanisms, the inversion framework provides a clean basis on which they can be tested using future data combinations.

The central limitation of the present analysis stems from the covariance of $X(z)$ with external cosmological parameters such as $H_0$ and $\Omega_m$.  
Because the inversion depends on the absolute normalization of the distance scale, residual degeneracies may suppress or exaggerate subtle curvature in $X(z)$.  
A consistent joint analysis combining BAO and cosmic–chronometer data will be necessary to mitigate this dependence and enhance the sensitivity to genuine evolution. 
In this regard, the approach presented here should be viewed as a baseline diagnostic—one capable of directly revealing departures from $\Lambda$CDM once higher–precision distance measurements from upcoming surveys such as DESI \cite{DESI:2016fyo}, LSST \cite{LSST:2008ijt}, or the Roman Space Telescope \cite{Spergel:2013tha} become available.


Overall, the discussion of our findings within the broader context of current cosmological constraints suggests that the data remain compatible with a constant dark energy density.  
Yet the method’s ability to isolate, visualize, and quantify possible departures establishes it as a valuable complement to parametric fits and likelihood analyses.  
Even if the small curvature seen here ultimately proves non–physical, the exercise highlights how inversion–based reconstructions can serve as an independent consistency check for the $\Lambda$CDM paradigm and as a sensitive probe for any genuine time variation that future data might reveal.

\section{Conclusions}
\label{sec:7}

The inversion analysis presented here provides a direct and fully empirical connection between the observed luminosity–distance relation and the underlying dark energy density function.  
By formulating the problem in terms of the derivative of the distance modulus, the method isolates the physical information encoded in the expansion history without assuming a prescribed functional form for $w(z)$ or $X(z)$.  
This approach establishes a complementary pathway to traditional likelihood analyses: one that is less sensitive to prior choices and capable of revealing departures from $\Lambda$CDM through the data themselves.

The reconstructed functions derived from the DES-SN5YR sample indicate that, within current observational precision, the energy density of the dark sector remains effectively constant.  
The analysis therefore reinforces the internal consistency of the $\Lambda$CDM model at late times, while simultaneously demonstrating the feasibility of differential, model–independent reconstructions on real data.  
Equally important, the absence of numerical instabilities or unphysical features confirms that the inversion technique can handle the statistical properties of large supernova datasets without distortion, providing a transparent diagnostic of data quality and error propagation.

Although no statistically significant evidence of dark energy evolution is detected, the framework developed here opens the possibility of revisiting this question with future high–precision surveys.  
Upcoming releases from DESI, together with Euclid, Vera Rubin, and Roman, will extend redshift coverage and improve calibration uniformity, reducing the degeneracy between $X(z)$ and external parameters such as $H_0$ and $\Omega_m$.  
Applied to these datasets, the inversion formalism will enable decisive tests of whether the apparent stability of $X(z)$ persists or conceals subtle departures from a cosmological constant.  

In summary, this work consolidates a rigorous, data–driven methodology for probing the dynamics of dark energy.  
Its strength lies not in detecting evolution where none is yet confirmed, but in establishing a foundation capable of revealing it unambiguously when the observational precision finally allows.

\section*{Acknowledgments}

JB would like to thank FIB-UV  for partial support.
VHC would like to thank CEFITEV-UV for partial support. 
CC acknowledges partial support by SNII

\bibliographystyle{apsrev4-1}

\bibliography{references}

\end{document}